\documentclass[12pt, preprint]{aastex}
\usepackage{emulateapj5}
\usepackage{epsfig}

\newcommand{\chandra}{{\it Chandra}} 
\newcommand{\xmm}{{\it XMM-Newton}}

\newcommand{\pulsar}{1E1207}
\newcommand\lax{\>\vcenter{\hbox{$<$\hskip-.75em\lower1.0ex\hbox{$\sim$}}}\>}
\newcommand\uax{\>\vcenter{\hbox{$>$\hskip-.75em\lower1.0ex\hbox{$\sim$}}}\>}

\begin{document} 

\title{Detailed spectral analysis of the 260 ksec \xmm\ data of 1E1207.4-5209 and significance of a 2.1 keV absorption feature} 
\author{Kaya Mori\altaffilmark{1}, James C. Chonko\altaffilmark{2} and Charles J. Hailey\altaffilmark{2}}
\altaffiltext{1}{Canadian Institute for Theoretical Astrophysics, 60 St. George St., Toronto, ON, Canada, M5S 3H8} 
\altaffiltext{2}{Columbia Astrophysics Laboratory, 550 W. 120th St., New York, NY 10027}
\email{kaya@cita.utoronto.ca, chonko@astro.columbia.edu, chuckh@astro.columbia.edu}

\begin{abstract}

We have reanalyzed the 260 ksec XMM-Newton observation of
1E1207.4-5209. There are several significant improvements over previous
work. Firstly, a much broader range of physically plausible spectral
models was used. Secondly, we have utilized a more rigorous statistical
analysis. The standard F-distribution was not employed, but rather the
exact finite statistics F-distribution was determined by Monte-Carlo
simulations. This approach was motivated by the recent work of Protassov
et al. (2002) and Freeman et al. (1999). They demonstrated that the standard
F-distribution is not even asymptotically correct when applied to assess
the signficance of additional absorption features in a spectrum. With our
improved analysis we do not find a third and fourth spectral feature (Bignami et al. (2003), De Luca et 
al. (2004)) in 1E1207.4-5209, but only the two broad absorption features previously
reported (Sanwal et al. (2002), Hailey and Mori (2002)). Two additional
statistical tests, one line model dependent and the other line model independent
confirmed our modified F-test analysis. For all physically plausible continuum
models where the weak residuals are strong enough to fit, the residuals occur
at the instrument Au-M edge. As a sanity check we confirmed that the residuals
are consistent in strength and position with the instrument Au-M residuals 
observed in 3C273.

\end{abstract}



\section{Introduction}

The thermal emission from isolated neutron stars (INS) probes neutron star physics since it is uncontaminated
by emission from an accretion disk. In particular, many 
INS (age $\sim10^3$--$10^5$ yrs) are good candidates because they are hot
enough that the thermal emission is not obscured by non-thermal magnetospheric
emission. Lines in thermal spectra are an excellent tool
to probe the strong magnetic field and gravity of the INS.

Two broad absorption features were discovered in the INS 1E1207 by
\chandra\ \citep{sanwal02} and later confirmed by \xmm\
\citep{mereghetti02}. The discovery was followed by the detection of a
single spectral feature in the X-ray thermal spectra of three nearby INS
\citep{haberl03, vankerkwijk04, haberl04}. \pulsar\ still remains unique because it
shows more than one spectral feature as opposed to the other three INS which
only show a single absorption feature.

The interpretation of features in NS thermal spectra is
not straightforward because of effects such as magnetic field,
gravity and unknown surface composition. 
Soon after the discovery of features in the \pulsar\ spectrum
several different interpretations were proposed ranging from
Helium atomic lines at $B\sim10^{14}$ G \citep{sanwal02}, Iron atomic
lines at $B\sim10^{12}$ G \citep{mereghetti02}, Oxygen/Neon atomic
lines at $B\sim10^{12}$ G \citep{hailey02, mori03_1} and hydrogen molecule ion lines at $B\sim10^{14}$ G \citep{turbiner04}. Recently,
\citet{bignami03} (hereafter B03) proposed electron cyclotron lines 
based on the detection of two additional absorption
features at 2.1 and 2.8 keV in the 260 ksec \xmm\ observations. They
interpreted these four absorption lines as the fundamental and three 
harmonics of cyclotron lines. \citet{deluca04}
(hereafter DL04) re-analyzed the same \xmm\ data and also found that the 3rd
and 4th feature are significant.

In this paper we present our detailed analysis of the same 260
ksec \xmm\ observation and investigate the significance of the two
additional absorption features at 2.1 keV and 2.8 keV. 


\section{Data reduction methodology \label{sec_data}}

\xmm\ observed the INS \pulsar\ for two
orbits starting on August 4, 2002. The PN camera was operated with a
thin filter in the "small window" mode. This allows accurate timing
(6 ms time resolution) and minimizes pile up. The
two MOS cameras were operated in "full frame" mode with a thin
filter. Our data reduction followed the latest \xmm\ calibration document \citep{kirsch04} and we processed the data by running the pipeline directly from the ODF files with SAS version 5.4.1. \footnote{Since submission of the manuscript we have confirmed our results remain the same with the more recent SAS version 6.0} We used
canned response matrices for all the instruments and generated
ancillary files by the SAS command "arfgen". We also generated response function using the SAS command ``rmfgen'', but the difference in chi-squared and fit parameters was tiny. Following \citet{kirsch04}, we generated a lightcurve of high energy ($> 10$ keV) single pixel (PATTERN = 0) events on the full field of view and selected good time intervals using the criteria for count rates $< 1.0$ cts/sec (PN) and $0.35$ cts/sec (MOS).   The total exposure time 
for PN and MOS after removing time intervals with high background 
are 156 ksec and 231 ksec respectively. 

We extracted the source photons from a 45" radius circle centered on
the source for all cameras. There are sufficient photons in the PN
camera to enable selection of only single events to achieve the best
spectral resolution. Most of our results are based on this higher quality
singles data, however later we discuss some results from the doubles
analysis. The MOS cameras have less photons so we used all
events in our analysis to improve the photon statistics.  

Background regions were selected according to the
recommendations of the \xmm\ calibration team \citep{kirsch04}. In the PN camera we chose a 45" radius circle
background region the same distance from the readout node as the
source region to ensure similar noise. An annular background region
encircling the source was not used to avoid out-of-time events from
the source.  In the MOS cameras we used a circular region away
from the source according to the calibration team recommendations.
We also tried an annular region
surrounding the source and found no significant deviations in the
continuum fit parameters. The source is located in a supernova remnant
and non-uniform X-ray emission from the remnant could influence the
neutron star spectra. To validate our choice of background region we
chose three different background regions for each instrument and
confirmed the spectral fit parameters of the source did not
significantly change with our choices. As a result, source count rates in 0.2--5.0 keV are 1.546+/-0.003 cts/sec (PN single+double), 0.346+/-0.001  cts/sec (MOS1) and 0.353+/-0.001  cts/sec (MOS2). Background count rates in the energy band are 0.089+/-0.001 cts/sec (PN single+double), 0.0065+/-0.0005  cts/sec (MOS1) and 0.0069+/-0.0005  cts/sec (MOS2). 

The spectra were rebinned such that each bin contained at least 40 
counts and analyzed using XSPEC v11.3.1 in the 0.3--4.0 keV energy 
range. Modest oversampling of the energy resolution kernel was utilized.
Our number of bins and degrees of freedom are comparable to those of
other analyses of this source. 
We adopted the energy band between 0.3 and 4.0 keV for the 
PN instrument. For the MOS instruments we selected the 0.5--4.0 keV band
because the MOS is not well-calibrated at lower energies. Continuum fit
parameters for the PN and MOS instruments were in good agreement. 

The long \xmm\ observation permits splitting the full observation
in order to search for time variability.  We split the observation 
into four approximately equal time intervals and found no deviations 
between the four observations. They yielded statistically 
consistent continuum fit parameters. 
We conclude that there was no time dependence during 
the total observation and therefore we use the data from the entire 
observation for all subsequent analysis.


\section{Analysis of data using continuum models \label{sec_cont}}

\subsection{Introduction}

In this section we consider fits to the global \pulsar\ spectra using a
variety of physically motivated spectral models. Most of these models
are not compatible with the data and are rejected. For those models
which provide good fits to the data we determine the relevant fitting
parameters. The issue of properly fitting additional
spectral features is a complex one, and we postpone addressing it
until \S\ref{sec_stat}. Here we only set the stage by providing information
on the best fit continuum models.

\subsection{Atmosphere models for neutron star thermal spectra 
\label{sec_atmos}}

The fitted continuum models must reflect the full range of plausible
INS atmospheric conditions. The surface of an INS can be any element
from Hydrogen to Iron, since only $\sim10^{-19} M_\odot$ of material
produces an optically thick atmosphere \citep{romani87}.  The
atmosphere distorts the spectral energy
distribution (SED) from Planckian, as will the presence of a
magnetic field. For $B=0$, hydrogen atmosphere models have the hardest
spectrum due to free-free absorption. INS spectra soften and show a
proliferation of lines and edges at higher energies, becoming closer
to black-body \citep{romani87}. For increasing B and atomic number there
is significant spectral softening of the SED \citep{ho02}. Even for Hydrogen,
electron binding effects can amount to an appreciable fraction of the INS
surface kT at high enough B-field, leading to incomplete ionization and
softening of the SED due to opacity from bound species \citep{ho03}. From the
above considerations we conclude that realistic atmospheric models for
\pulsar\ will have an SED somewhere between black-body and a fully-ionized
hydrogen model, and our subsequent fitting reflects this range of SED.

\subsection{Description of model fits to the data (no third or fourth line 
assumed) \label{sec_model}}

In the following few sections we describe our fit to the data without
the assumption of a third or fourth line in the spectrum. Fits with
such additional lines are discussed in \S\ref{sec_stat}.  Therefore our
"baseline" models always consist of the two universally accepted lines
in the \pulsar\ spectrum, which we model as Gaussian absorption lines at
0.7 and 1.4 keV.

For each continuum component we adopt one of the following three 
physically plausible models; blackbody (BB), magnetized hydrogen 
atmosphere (HA) or power-law (PL).  The magnetized hydrogen atmosphere
model was constructed assuming a fully-ionized hydrogen atmosphere at
$B=10^{12}$ G, close to the dipole field strength as measured from the
spin-down parameters (private communication with V.E. Zavlin). We also tried B-fields as low as $B=0$ \citep{zavlin96}, and
this did not alter the continuum parameters significantly. Higher B-fields ($\uax 4.4\times10^{13}$ G), just give results closer to the  black-body case \citep{ho02}. 

In our first fits to the spectrum we used one continuum component
along with the two absorption lines. For all instruments the
one-component models did not yield acceptable $\chi^2$ values
($\chi^2_\nu >$ 1.3) and showed significant residuals above $\sim 2$
keV.  This suggested the need for another continuum component.

Given two continuum components and two absorption lines there are
three types of continuum models defined as below.  They consist of two
continuum components ($C1$ and $C2$) and two absorption lines ($L1$ and $L2$).
$C1$/$L1$ and $C2$/$L2$ refer to the lower and higher energy spectral
component respectively.  The three classes of models considered (with
the notation $C*L$ meaning line $L$ resides on continuum component $C$)
\begin{eqnarray}
\mbox{Model I:} && C1*L1*L2+C2\nonumber\\
\mbox{Model II:} && C1*L1+C2*L2 \nonumber \\
\mbox{Model III:} && (C1+C2)*L1*L2 \nonumber
\end{eqnarray}
For $L1$ and $L2$ the Gaussian absorption line was modeled as
\begin{equation}
F(E) \propto \exp\{-\tau \exp(-(E-E_0)^2/2w^2)\}
\end{equation}
where $\tau$ and $w$ refer to the line depth and width.  We fit
other absorption line profiles such as a Lorentzian but the results
were similar. While the exact shape of the absorption features may be
asymmetric or have substructure from blended lines \citep{mori03_1},
we obtained excellent fits without invoking asymmetric profiles. Thus 
we chose simple Gaussian line shapes. When we fit photo-absorption 
edges to the two absorption
features at 0.7 and 1.4 keV, the $\chi^2_\nu$ was not as
good as for the Gaussian lines in any of the continuum models.

\subsubsection{Physical meaning of model I, II and III \label{sec_physics}}

Interpretation of spectral features in NS thermal spectra is not
straightforward since the line parameters depend on various NS
parameters such as surface element, ionization state, magnetic field
strength and gravitational redshift \citep{hailey02,
mori03_1}. Therefore, it is important whether the observed spectral
features originate from the same region or layer.

The presence of two continuum components complicates our
interpretation of the observed absorption features. We assume that the
two continuum components originate from different regions on the
NS surface. For instance, $C1$ is emitted from a large area on the
surface and $C2$ is emitted from a hot polar cap (DL04). 

Model I assumes that $L1$ and $L2$ are from the same region as the
emission of C1 and Model II assumes that they are from different
regions. Model III assumes that $L1$ and $L2$ are from a layer above
the two regions emitting continuum photons. Several physical models
predicting a layer made of electron-positron or electron-ion pairs 
a few NS radii above the NS surface have been proposed
\citep{dermer91, wang98, ruderman03}. Such a layer may become
optically thick and modify the thermal spectra from the NS surface. In
addition, absorption in model III can take place anywhere between the
NS and the observer (e.g. magnetosphere, supernova remnant, ISM or
materials on the \xmm\ mirrors and cameras).

\subsection{Results of continuum model fitting to \xmm\ data \label{sec_fit}}

We searched for continuum models that are consistent with the data
using models I, II and III. The set of models we considered is extensive
including all
permutations of blackbody (BB), magnetized hydrogen atmosphere (HA)
and power law (PL) models for $C1$ and $C2$.

All the two-component thermal models (and combinations of BB and HA)
fit the data well, yielding $\chi^2_\nu$ very close to unity (table 1 and
2). For brevity we do not consider mixed cases (BB+HA) because tests
showed that in assessing absorption line significances (\S 4) they
always produced results intermediate between the pure BB and pure HA
cases.  Continuum models with PL components are ruled out because they
do not adequately fit the data, leaving significant residuals above 2
keV.  Models I, II and III with thermal continuum components are all
acceptable. 

\begin{figure*}[ht]
\epsscale{2.0}
\plotone{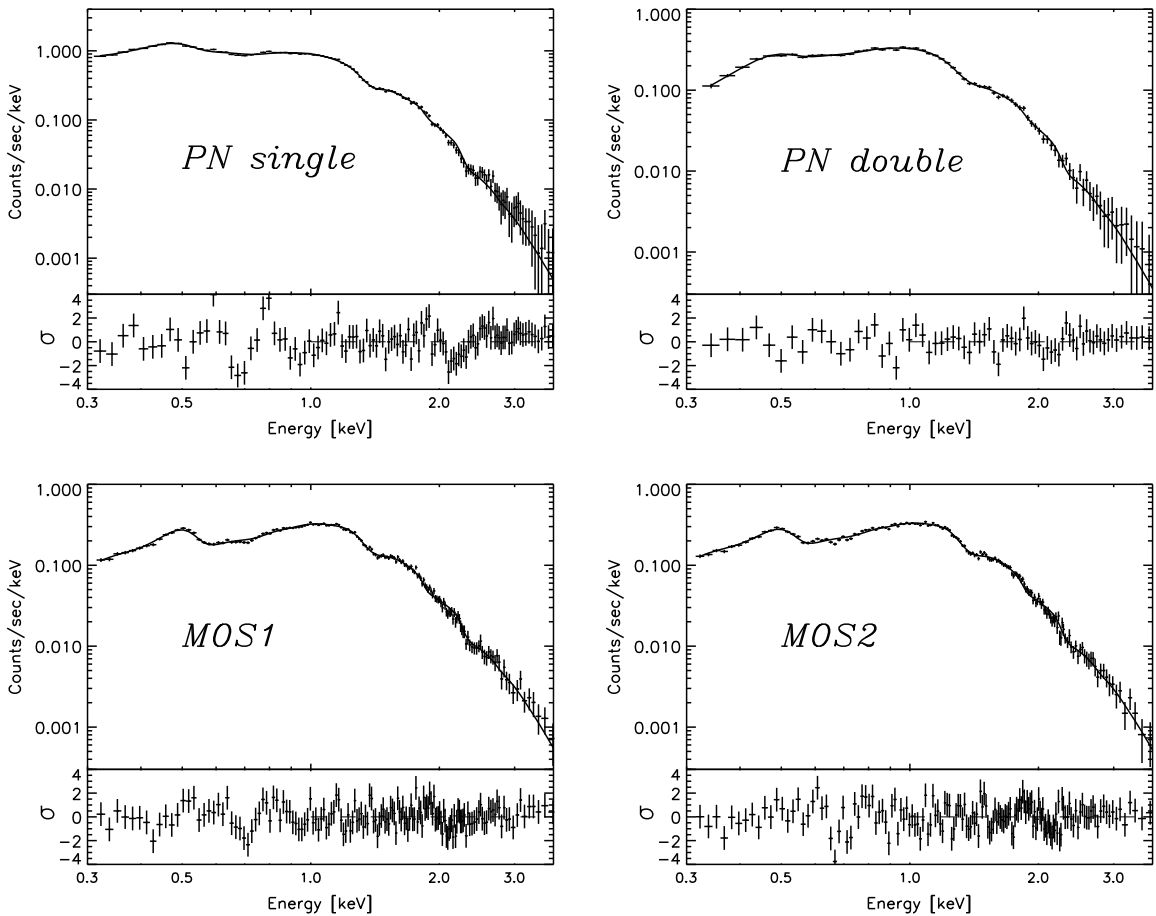}
\caption{PN (top) and MOS (bottom) spectra fit by I(BB+BB) 
model and their residuals. The excess around 0.7 keV is a known calibration feature \citep{kirsch04}. \label{fig_spec}}
\end{figure*}

Figure \ref{fig_spec} show the spectra and 
residuals for the best fit to
model I(BB+BB) using the PN single/double events and MOS1/MOS2 data. We also fit the data with a third 
spectral line, calculated $\chi^2$, and
evaluated the difference of $\chi^2$ between third line case and no third line case.
This is a key parameter in our subsequent statistical tests. 

We show our best-fit continuum parameters and 90\% confidence levels
for six different models in the energy range from 0.3-4.0 keV for PN
in Table \ref{tab_pn_fit}. All the errors quoted are statistical errors and we did not include any systematic error. We calculated 90\% confidence level errors using $S=S_{min}+2.706$ for each parameter while allowing the other parameters to vary freely \citep{lampton76}. These models yield excellent fits to the data with $\chi^2_\nu$ values for PN slightly less than unity.  We found that
Gaussian line profiles yield a statistically good fit to the data.
The PN blackbody models I, II and III all give consistent values for
these parameters. 

We note that for some of our models the line depths (especially for the 1.4 
keV line) 
are not well-constrained.  This is caused by
the presence of two continuum components. If the absorption
feature is fixed to reside on one of the continuum components then the
level of that component becomes very important.  If the depth of the 
absorption
feature is close to the continuum level the residuals cannot be
adequately fit rendering the line depth poorly constrained. It 
is possible for both of
the broad absorption features to lie on the higher temperature continuum
component $(C1+C2*L1*L2)$. However, the fit was not acceptable leaving 
significant deviations from the data.

As an aside we note that we took an additional step to confirm the 
robustness of our continuum parameter fitting and $\chi^2$ determination.
We examined a range of degrees of freedom in the PN analysis spanning those
previously reported in the literature, and consistent with obtaining normal
statistics and sensible energy resolution kernel sampling. Both our continuum
parameter determination and quality of fit were insensate to this. However we
emphasize the extremely important technical point that we selected the actual
bin size a priori, not by adjusting the size emipirically to obtain the lowest
$\chi^2$. It is well known this would result in a fit statistic which is
not $\chi^2$ distributed (Eadie et al. 1983). It is important for the 
subsequent statistical analysis that the conditions for a distribution-free
statistic are met, even in the finite-N limit.

\subsection{Comparison with and improvements to previous work \label{sec_comp}}

In several previous investigations (DL04, B03) the data were fit with two
black-body components and two absorption lines. In both cases $\chi^2_\nu$
significantly greater than one were obtained, requiring the inclusion of
additional absorption lines to fit some higher energy residuals. Other work
did not require such extra lines when fitting either black-body continuua
\citep{sanwal02, hailey02, mereghetti02} or magnetized HA models \citep{mereghetti02}. 

In order to more thoroughly investigate these deviations in fit quality
using just continuum models we have exploited, in addition to more up-to-date
response matrices and a more time-consuming study of background effects,
a much broader set of continuum models (models I and II in addition to model
III). In the case of model III, where our comparison with previous work is
direct, we find agreement at the 90$\%$ confidence level for all derived parameters
for both MOS and PN data. 

In regards the overall fits, we find acceptable $\chi^2_\nu$($\sim$ 1--1.3)
for all three classes of models, i.e., our fits do not
require additional spectral features. However introduction of extra lines
does improve the overall $\chi^2$ very slightly. Assessing the necessity
of such additional features is a matter of great statistical delicacy, which
we discuss fully in \S 4. 

To facilitate a direct comparison with previous work we have used the results
of table 2 in DL04 which indicates the best continuum model fits (without 2
additional lines) as giving $\chi^2_\nu \sim$1.8, 1.3 and 1.6 for PN, MOS1 and
MOS2 respectively. While the MOS fits of DL04 are not as good as ours (for any
of our three models), the discrepancy is not really that large; DL04 only 
marginally required extra lines in the MOS data. Only the PN fits appear to have
a large "discrepancy". One must be very cautious in assuming this discrepancy
arises from differences in processing of the data. Such an assumption is belied
by the agreement in overall signal and background count rates in MOS and PN,
in MOS and PN fitting parameters and even in overall $\chi^2$ for MOS1/2 data
between all groups. The problem is that in the case of finite count statistics,
there can be substantial deviations in $\chi^2$ from its ideal, distribution-free
theoretical form \citep{eadie}.  

Simulations support this assertion. We generated 1000 sets of bootstrapped data 
based on the 1E1207 data set, fitting this data set with continuum model III, and
then comparing the theoretical and observed $\chi^2$ distributions.  It was 
found that $\chi^2_\nu$ of 1.6 or greater were found more than an order of
magnitude more times than predicted by the standard $\chi^2$ distribution. Thus
there is a non-negligible probability ($\sim$ 10\%) of obtaining a deceivingly
large $\chi^2$ using proper fitting procedures. The difference in $\chi^2_\nu$
for the PN data may be nothing more than artifacts of finite N statistics. Our
results are more immune to this because we fit so many test cases. Moreover
such a large $\chi^2_\nu$ is not problematic per se. Rather what is required is
an appropriate finite-N statistical methodology applicable to each reduced
data set. An appropriate approach has recently been developed and we present it
and apply it to the 1E1207 data set in \S 4 and 5.


\section{Statistical tests for extra spectral features \label{sec_stat}}

While it may naively appear that assessing the existence of and need for an
extra spectral feature in a continuum spectrum is straightforward, this is not
true. We present three approaches for assessing the need for extra spectral
features. These include goodness-of-fit test for lines in a continuum, 
direct spectral fitting and significance assessment, and the modified F-test.

We have chosen to apply the F-test because it has been used in previous work
on 1E1207. Despite its great familiarity to astrophysicists, the use of the
F-test to determine the need for extra spectral features is extremely 
problematic. To effect this analysis we exploit the recent work of Protassov
et al. (2002) (hereafter P02) and Freeman et al. (1999) (hereafter F99). These authors
have developed approaches which surmount the difficulties in the standard
F-test and permit much more reliable and accurate determination of feature
significance than is possible with the "standard" F-test. Our analysis exactly
follows that of P02, so we emphasize P02 in our brief discussion of theoretical
and practical particulars in \S 4.1. The reader is directed to the above
references for a more thorough treatment.  In
\S\ref{sec_eadie} and \S\ref{sec_ew} we consider the other methods for
attacking this problem.


\subsection{F-test for detecting extra model components \label{sec_like}}

In the F-test and its closely related cousin the likelihood ratio test (LRT)
a test statistic is formed from the distribution function $f(x_i|\theta)$ 
governing the probability of counts $x_i$ in bin $i$ under a set of parameters
$\theta=(\theta_1....\theta_p)$. The test statistic is a functional containing the
$f(x_i|\theta)$ with $\theta$ free to assume any value and also $f(x_i|\theta)$ where
some parameters are constrained by the values that are required by the null
hypothesis. To be concrete, in the case of the F-statistic the statistic 
$F=(\chi^2_c(x,\theta)-\chi^2_u(x,\theta))/\chi^2_\nu$ (Bevington 1969; P02).

The F-test is extremely powerful because it is ``distribution-free''. That is,
in the asymptotic (large count) limit, and under the null hypothesis (specifically, 
in our case, $\tau=0$) it does not depend on the initial probability
distribution, which is the underlying distribution $f(x_i|\theta_1...\theta_p)$.  
Rather the test statistics only depend on the
reference distribution (in our case the F-distribution). The asymptotic
behavior effectively decouples details such as the shape of the continuum and
the particulars of the line profile and strength from the way the test statistic
is distributed asymptotically. This is why the test is so powerful and broadly
applicable. 

P02 and F99 made the crucial observation that there are certain
regularity and topological conditions that must be met for distribution-free
statistics. The regularity conditions are related to the integrals and derivatives
of $f(x_i|\theta_1...\theta_p)$ (see Serfling 1980; Roe 1992 for a precise statement) and
are of no interest to us since these conditions are weak and thus almost always
satisfied in astrophysics modeling. Of profound importance are the topological
conditions (P02). These are: 1) the model under test must be nested, that is, the
parameter under test in the constrained model must be a subset of that parameter
in the unconstrained model and 2) the parameter under test in the constrained 
model cannot lie on the boundary of the possible values of the parameter under
the unconstrained test. 

Evaluating these topological conditions for equation 1 we see that the first is met;
the constrained model has $\tau = 0$ and thus the parameterization in the constrained
model is a subset of that in the unconstrained model. However condition 2 is violated.
$\tau=0$ is on the boundary of the possible values $\tau$ can assume under the
unconstrained model. When condition 2 is violated {\it the F-statistic does not
even asymptotically approach the F-distribution}. P02 illustrate, through simulations
for an absorption line on a continuum, that the reference distribution overestimates
the strength of the line by almost an order of magnitude. P02 properly note that in
general nothing can be said about the significance of a line with the standard
F-test, because we do not even in principle know the asymptotic reference distribution.
In this regard, for instance, the XSPEC users manual specifically advises against
the use of the F-test in assessing the necessity for extra lines in a spectrum.

\subsubsection{Analysis of 1E1207 line significance using modified F-test 
\label{sec_mlrf}}

Following P02 we seek to more accurately assess the significance of the alleged
third and fourth lines in 1E1207 by employing the F-statistic, which we now
recognize is not F-distributed. The F-statistic is actually a less obvious 
choice than LRT, but it facilitates direct comparison with previous work. Since
an appropriate asymptotic distribution does not exist, even in principle, what is
now required is the establishment of a reference distribution for finite statistics.
P02 and F99 explain in detail how to do this. We have determined the
reference distribution using the posterior predictive p-value methodology (PPPM)
of P02, and the reader is referred there for details. The idea behind PPPM is
straightforward although the computer resources required to implement it are 
substantial. We only outline the procedure here.

Under the null hypothesis we can use data to generate fake data from which we
can determine the relevant reference distribution.  However this standard 
bootstrapping technique is not correct, since we do not know the "true" values
of the parameters to use in the simulation. The failure of the topological
condition means we lose conditioning properties which would make it meaningful to
replace parameters by best estimators. The solution, as pointed out by P02, is to
use parameter values in the simulation which are likely given the observed data.
This simply requires a statistical distribution of parameter values which can be
sampled and incorporated into the simulation which generates the distribution of 
the statistic. P02 advocate a Bayesian approach for determining the probability
distribution of each parameter, given the data. PPPM is a specific Bayesian
implementation. In effect, it is a fancy statistical bootstrap with a method for
explicitly incorporating parameter fitting errors. The entire approach is completely
straighforward to implement in a Monte-Carlo by brute force, althouth clever and
efficient non-brute force methods for implementation of the Monte-Carlo exist \citep{vandyk01}.


\subsubsection{Details of the line search procedure}

We simulated our spectra with 
parameters determined by Bayes formula. These spectra are folded
through the instrumental response using XSPEC. The actual line search
was done by performing a blind search for absorption lines between 0.3
and 4.0 keV. The statistic $F$ was calculated as, 
  $F = (\chi^2_c-\chi^2_u)/\chi^2_\nu$ (\citet{bevington69}; P02) where $F$ is obtained
from the fit to a continuum model (C) and continuum model plus
absorption line (U).  A blind search for spectral features in fake
spectra in XSPEC is dependent on the initial values and paths 
\citep{rutledge02}. In order to reduce this effect we started
the line search from four different line energies. We may not find all
the significant absorption lines when more than four are present in
the simulated spectrum. However such cases, which entail a large F
statistic, occur very rarely. To confirm this we performed Monte-Carlo
simulations with an increased number of initial starting energies, and
found that the number of occurrences of $F$ larger than 5 did not
change. In searching for a line from four starting locations we found
that sometimes XSPEC fit the same feature more than once. We corrected
for such double-counted events {\it a posteriori}.

\begin{figure*}[ht]
\epsscale{1.8}
\plotone{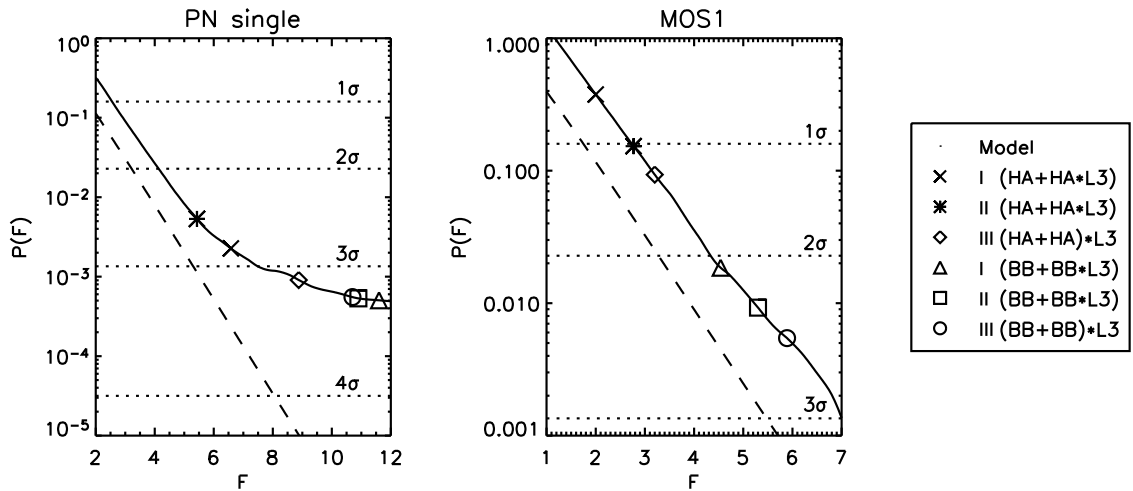}
\caption{$P(F)$ distribution for PN-single (left) and MOS1 (right) instrument. The dashed line shows the classical F-distribution. \label{fig_fdist}}
\end{figure*}

The distribution of the statistic $F = \Delta\chi^2/\chi^2_\nu$ is shown in
figure \ref{fig_fdist}. This distribution was determined for each instrument. In
formulating this distribution we recall \S\ref{sec_like} that this 
corresponds 
to the distribution of $F$ under the null hypothesis, i.e., that there is
no line.  This is what is shown in figure \ref{fig_fdist}, along with the classical
F-distribution. The value of F for six models is plotted, along with significance line. We conclude that there are no grounds for rejecting the null
hypothesis (no third line). The results are always in the 1--3$\sigma$ range and
only reach the higher level in a few isolated instances involving 
black-body models and PN singles data.

Our results differ from previous work (B03/DL04) for two reasons. Firstly our 
F-statistic 
is about 50\% smaller due to our better fits to the continuum. And secondly 
the use
of the correct finite distribution for the F-statistic drops the
significance of an extra absorption line by a huge amount, as happened in 
the likelihood
simulations of P02. The net result is a more than six order of magnitude 
reduction in the significance of the strongest line found in the PN data.

Further support for our conclusions is presented below, where
essentially the same result is arrived at by indepedent statistical tests.

\subsection{Goodness-of-Fit test for lines in smooth spectra \label{sec_eadie}}

This type of test is described in numerous papers, but 
we mainly follow the arguments and notation of \citet{eadie}.
We test the null hypothesis ($H_0$) that there is no deviation of counts with
respect to the continuum level against the alternative ($H_1$) that there is
such a deviation. We define a test statistic $T$, and unless $T$ exceeds some
critical
value $T_c$ then we accept $H_0$. Under $H_0$, $T$ has some distribution $P(T)$.
$T_c$ 
can then be implicitly defined by the formula $\alpha = \int_{T_c}^{\infty}
P(T) dT$
where $\alpha$ is defined by the 4$\sigma$ tail of a normal, standard
distribution.
This procedure ensures that the null hypothesis will not be rejected unless a
$4 \sigma$ burden of proof is met. This test is binary - unless $T$ exceeds
$T_c$ we 
accept $H_0$. This result can be further quantified by calculating
$1-C = \int_{-\infty}^{T} P(x) dx$. $1-C$ can be converted into units of
$\sigma$ using
the same method as employed for $\alpha$ and provides a rough measure of the significance of the absorption feature. 

We need only define our test statistic $T$ and its probability distribution
under
$H_0$. We are searching for a deficit of counts in a region of the spectrum
beyond that predicted by our continuum model. We will assume that the region 
of interest is defined by $K$ energy bins around 2.1 keV.
Define $n_i$ = observed counts in bin $i$; $s_i =
n_i-\hat{b_i}$, $b_i$ = true continuum in bin $i$ and $\hat{b_i}$ = estimated
continuum in bin $i$. We can then formally construct the null hypothesis
$H_0$, there is no line, ($s_i = 0$ for all $K$) against all alternatives
defined 
by $H_1$. This is the classic goodness-of-fit test.
The natural test statistic is $T=\sum_{i=1}^{K} \frac{(n_i-\hat{b_i})^2}{var(n_i-b_i)}$.

This can be expressed, under suitable conditions (appendix) as
\begin{equation}
T=\sum_{i=1}^{K}\frac{(n_i-b_i)^2}{\hat{b_i}+\hat{\sigma_i}^2}.
\end{equation}
This test statistic is composed of random variables $z_i = (n_i-
b_i)/\sqrt{(n_i-\hat{b_i})}$ or equivalently $T = \sum z_i^2$.  We restrict
this sum to the $K$ bins of interest. Under the null hypothesis $H_0$
one can show (appendix) that for $n_i$ not too small that the $z_i$ are
normal
(Gaussian) and standard (mean 0 and variance 1). Thus under $H_0$, $T$ is
by construction $\chi^2$ distributed with $K$ dof. If $K \uax 30$ one can
additionally show (appendix) that a new variable can be constructed
$y = (T- K)/\sqrt{2K}$ and $y$ is asymptotically standard,
normal. We select a $T_c (y_c)$, and if $T < T_c$ $(y < y_c)$ we accept $H_0$
(there are no counts in deficit of the continuum model). $T_c$ is related to
$\alpha$
as discussed above. $y_c$ is similarly defined only with the standard, normal
distribution replacing $P(T)$.

\subsubsection{Results applied to 1E1207 data}

 For all the models we considered the observed $T$ was much less
than $T_c$
so we accept $H_0$ - i.e., there is no line. Table \ref{tab_eadie} lists the observed
$T$ and $T_c$ for various models. Note $T_c$ changes for each model because of
its dependence on the continuum level and fit variances within that model.
Table \ref{tab_eadie} also indicates $1-C$, to give some feel for the "distance" of $T$ from $T_c$ in
probability space. Because $K$ is sufficiently large in this analysis, we formed
the random variable $y$ from the observed $T$, as mentioned above. This is shown in
table \ref{tab_master} and provides a direct comparison of how close (far) the measured $y$ is
from $y_c$ and the "center" of the distribution $P(y)$ in units of $\sigma$.

The formulation above is correct in searching a region where there
is {\it a priori} reason to expect a line. There is a complication in how
exactly to determine the $K$ bins of interest (other than that they are
around 2.1 keV). If the spectral feature is narrow
and the region of the $K$ bins broad, then the feature can be washed
out. This is also true in the converse case. To be conservative we checked
varying sized regions of 
interest around 2.1 keV. The results were not significantly changed
for regions comparable to a few times the energy resolution (from less than to
slightly greater than the widths reported in B03/DL04).

In such a blind search one must correct for the fact that there are N 
possible regions
of interest ($N$ is the energy band divided by the width of a single region of
interest) in which to find a statistical fluctuation of a given level. Very
roughly this increases $1-C$ by the same factor, which is $\sim6$. A more detailed 
treatment 
can determine the exact factor \citep{eadie}.  We have neither corrected for 
this factor
in this section nor reduced the significances in the next section 
accordingly. Thus, it should be noted that the actual significances are approximately 6 times less than indicated in table 3. 


\subsection{Significance of the absorption lines by equivalent width \label{sec_ew}}

\begin{figure*}[ht]
\epsscale{1.0}
\plotone{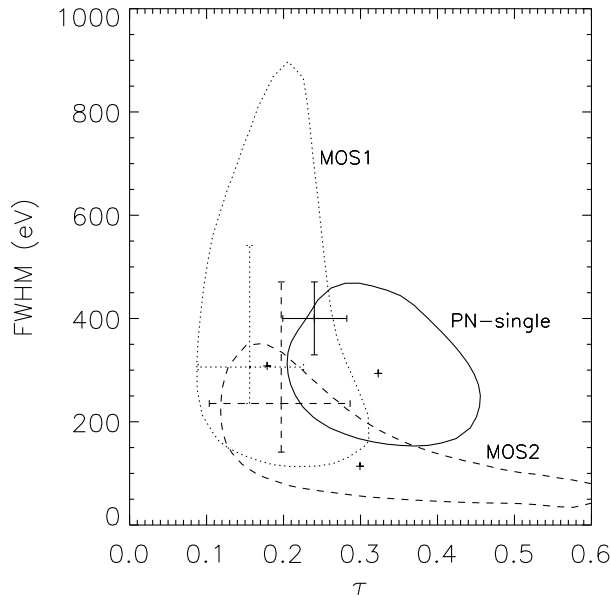}
\caption{$\tau$--$w$ contour plot for a 3rd line (PN, MOS1 and MOS2). The
contours were calculated at 68\%
confidence level. We fit model III(BB+BB)*L3 to the data. The crosses show the best-fit parameters at 90\% confidence of DL04 (Solid = PN, dashed = MOS1, dotted = MOS2). \label{fig_cont_all}}
\end{figure*}

The final and perhaps most straightforward method for deducing an
absorption feature's existence is to simply assume it exists and
estimate the significance of the detection.  In this section we
evaluate the significance of the residuals around 2.1 keV by fitting a
Gaussian absorption profile, and then calculate the equivalent width
(EW) and error bars. This particular absorption line profile gives
excellent fits to the data ($\chi^2_\nu\sim1$), so no improvement would be
had by alternate line shapes. We confirmed other line shapes did not
statistically significantly affect either the EW or the line
position. Of critical importance here is a proper calculation of the
error on the line fit. This is a standard non-linear $\chi^2$
fitting problem of the type discussed by \citet{lampton76} (LMB).
We follow their approach exactly. The 
spectrum is fit allowing all the continuum parameters to vary freely,
and the line depth and width are stepped through. The minimum value of
$\chi^2$ is jointly estimated for the line depth and width. A
contour plot of the 68\% confidence interval for line depth and width
is  then calculated using the prescription of LMB. Using the 68\%
$(\tau,w)$ contour we integrate to find the EW extrema, and the best fit
$(\tau,w)$ provide the best estimator of EW. We held the line centroid at
its best-fit value during this procedure to save computational time. A
series of test runs indicated that the EW depended only weakly on line
centroid within the 68\% confidence interval. The contour plots (68\%
confidence) are shown for the PN single and MOS1/MOS2 data in figure
\ref{fig_cont_all}, for model III(BB+BB)*L3.  In the case of the MOS data the errors in
depth and width are highly correlated, highlighting the importance of
a joint estimation of the errors using the procedure of LMB.

\begin{figure*}[ht]
\epsscale{1.2}
\plotone{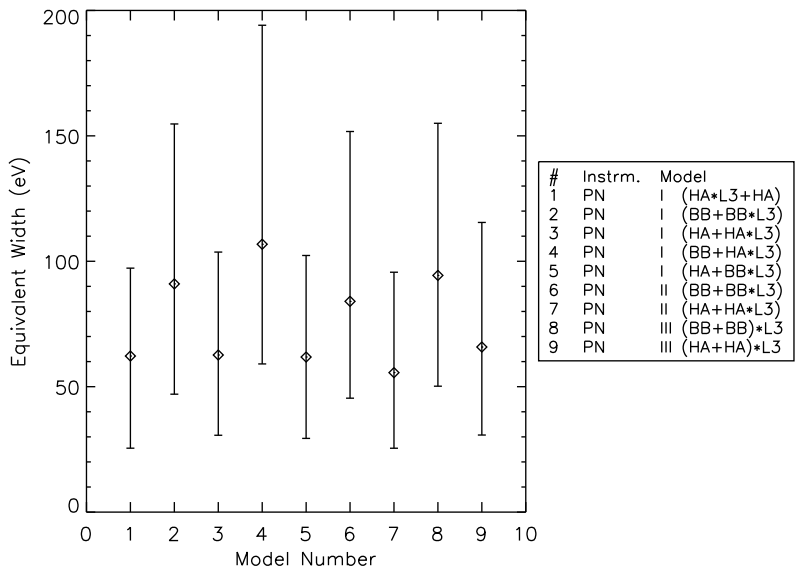}
\caption{Equivalent width [eV] of a 3rd line at 2.1 keV and 68\% confidence
levels. \label{fig_ew}}
\end{figure*}

The results of this analysis are shown in table \ref{tab_master} for all the relevant 
models. In no case is the statistical significance of the line detection larger than  $\sim2\sigma$. As noted in \S 4.2, this is a firm upper limit, uncorrected for a blind search. The strongest residuals are for PN data, and the EW for these cases are shown in figure \ref{fig_ew}. Except for the black-body models, the MOS1/2 residuals were essentially non-existent, so we gave up attempting to fit them. In some cases in table \ref{tab_master} the absorption line resided on a low temperature black-body
component which was so weak that no EW could be determined.  We note for 
thoroughness (although we think it obvious) that one must calculate 
the EW using both the low and the high energy continuum components, 
even though the physical model may indicate the line only resides on 
one continuum component. Calculating the EW in this latter case may be 
physically important in interpreting the physics, but in assessing the 
existence of a line only the former procedure has statistical 
significance.

\subsection{Concluding remarks on line search}

We have used three different methods to demonstrate there is no spectral
line in the 2.1 keV region. A summary is presented in table \ref{tab_master}. The level
of significance of the third spectral feature and measures of its significance
are consistent with each other. In only a few isolated cases of black-body 
continuua does the line rise even to the $3\sigma$ level (and only for the F-test), much less something
firmly indicative of a line. Since the fourth spectral feature is much, much
weaker than the third feature, we were unable to analyze it in detail. In \S 6
below we will present plausibility arguments that the statistically 
insignificant 
spectral residuals in the PN data are due to instrumental effects.

\section{Phase-resolved spectral analysis \label{sec_phase}}

\begin{figure*}[ht]
\epsscale{1.5}
\plotone{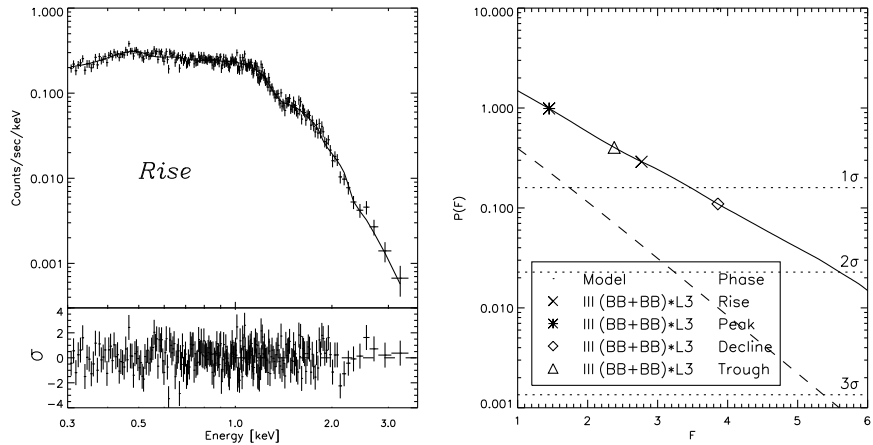}
\caption{One of the four phase-resolved spectra and residuals fit by model I (BB+BB) (left) and $P(F)$ distribution for phase-resolved spectra (right). The dashed line shows the classical F-distribution. \label{fig_phase}}
\end{figure*}

While no evidence of a third spectral feature was found in the
time-averaged data, for thoroughness we decided to repeat our analyses
for phase-resolved data. First we searched for the spin-period using
the PN data after correcting photon arrival times to the
barycenter. PN was operated in small window mode with time resolution
$\sim 6$ ms. The MOS data does not have sufficient time resolution for
spin-period determination. We used XRONOS 5.20 for our timing
analysis. Using the Epoch-folding method we found significant
pulsation at $P=424.13076\pm0.00002$ msec, consistent with the results of B03
and DL04. After tagging photons with the best-fit spin-period, we
reduced spectra in four spin phases. Spectral
bins were binned to have at least 40 counts in each bin. Figure
\ref{fig_phase} shows a typical phase-resolved spectrum fitted with continuum
model I(BB+BB) and residuals. We fixed $N_H$ to the fitted value
($1.32\times10^{20}$ cm$^2$ from the phase-averaged PN spectral
analysis).  Our results did not change when we let $N_H$ vary in
spectral fitting.

Around 2.1 keV, only a few spectral bins deviate from the continuum
model, and residuals are comparable to 
or smaller than the energy resolution, indicating that they are due to 
statistical fluctuation. This is borne out by statistical analysis of the 
residuals using the approach of \S\ref{sec_like}. Figure \ref{fig_phase} summarizes the results for the statistical tests on the phase-resolved data. There are no spectral features around 2.1 keV.  

\section{Physical significance of the residuals in the continuum spectrum}

We have demonstrated that there is no third line in the \pulsar\ spectrum.
Thus any attempt  
to explain the origin of $\lax$ 1--3 $\sigma$ residuals should be considered
suspect. We have several ground rules in this section. We explicitly avoid
detailed fitting of line shapes and in-depth searches for specific instrumental origins of the residual. It would certainly be futile given the strength of the residuals, and
we are unlikely to offer deep insights into possibly subtle instrumental
effects. Finally we concentrate on motivating our suspicions using PN
data, since the residuals are marginally stronger and thus likely to be more
fruitful hunting ground.

\subsection{Analysis of residual line centroids}

The analysis of \S\ref{sec_ew} yields the centroid of the weak residuals
for each of the models described in \S\ref{sec_cont}. A plausible hypothesis
is that the residuals are associated with the Au--M edge, which
occurs in the PN and MOS instruments. To pursue this hypothesis we need to
focus particular attention on the line centroids associated with model
III. If the residuals arise in the Au--M5 edge of the instrument, then
model III clearly corresponds to the only physically correct model for
the data. So if this hypothesis is true, we expect to see residuals around
the Au--M5 energy.

\begin{figure*}[ht]
\epsscale{1.0}
\plotone{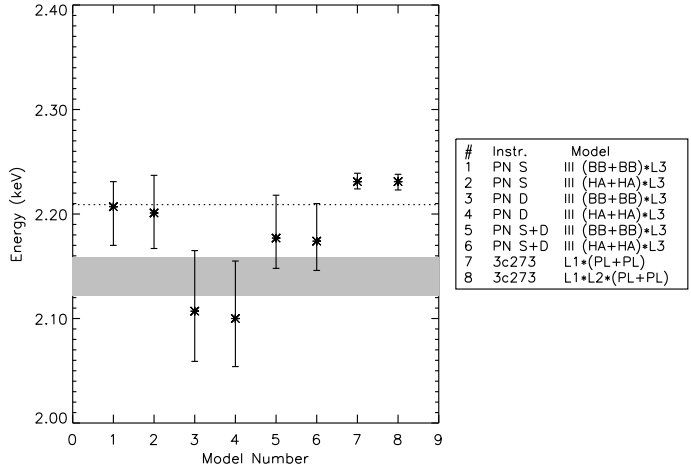}
\caption{Line energy of a fitted 3rd line.  The shaded region indicates the
90\% confidence level found by DL04. All error bars are 68\% confidence. The dotted line is at Au--M5 edge position. \label{fig_energy}}
\end{figure*}

Figure \ref{fig_energy} shows the line centroids from fits to the third line
using model III for PN single data, PN double data and combined PN single and
double data. We also show the
B03/DL04 90\% confidence interval band for their line centroid
determination. The dotted line is the theoretical position of the Au--M5 edge.
The striking feature of this plot is that the PN single data, which is
without
question the most reliable data since it rejects split events, occurs at
the position of the Au--M5 edge. Moreover the potential origin of the lower
residual
energy found in previous work is suggested by the fact that the PN doubles (split
events) residuals
occur at lower energy. Our combined PN singles/doubles data set is in agreement with
the results of B03/DL04. Thus we conclude the PN split events have reduced 
the apparent energy of the residuals,
while the pristine PN singles events are at the expected energy for an
instrumental effect. One would be tempted to say this lower energy in the
split
events is expected. This would get into
a discussion about how well split events are reconstructed in EPIC data and 
is far too detailed for us. We simply note that the highest quality event
data
shows residuals at the Au--M5 edge and the lowest quality data shows it at
much
lower energy. We also show in figure \ref{fig_energy}
the position of the residuals in 3C273, using a fitting procedure described
in the
next section. The 3C273 residuals, which are most certainly associated with the
Au--M5 edge, define the "fit" position of the Au--M5 edge and is a more
relevant
metric than the theoretical position of the edge. Nevertheless, the position of the Au--M feature in 3C273 is consistent at 
the 90\% level with the theoretical Au--M energy and it is also consistent with the 
position of the residuals 
in the PN singles data. A better line shape approximation would no doubt 
move the 3C273 result downwards, but this would require detailed understanding of 
the instrument
response. Our results are meant only to be suggestive since, after all, the
residuals are statistically insignificant.

\subsection{Consistency test between PN and MOS data \label{sec_consis}}

In DL04 a large inconsistency ($\sim 5\sigma$) between the significance of the lines
detected in PN and MOS data was reported, despite fit parameters being in 
statistical agreement. They explained this as being due to low energy calibration
uncertainties in the MOS. This observation motivated us to perform a consistency check on our data set. Because we utilized a MOS
energy cut slightly higher in energy, one would reduce low energy
calibration uncertainties, and thus better consistency between MOS and PN data sets.
This is what we have found, as described below. 

\begin{figure*}[ht]
\epsscale{1.5}
\plotone{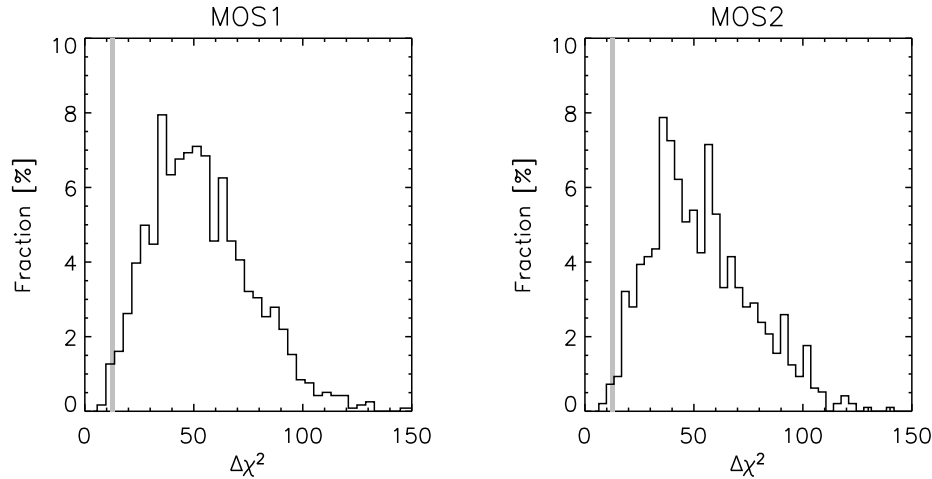}
\caption{Results from the consistency test for PN-single data. The
grey lines correspond to the measured $\Delta\chi^2$ values in MOS1
and MOS2 data. \label{fig_compare_pns}}
\end{figure*}

We set up a Monte-Carlo simulation to test whether the 3rd line residual
detected by PN could
have appeared in MOS data as a deficit with a significance comparable to that
observed. We used the III(BB+BB)*L3 
model fit to the data because it showed the 
largest significance of a 3rd feature, therefore the largest discrepancy between PN
and MOS. Hereafter
we describe our formalism using PN-single and MOS1 case as an example:
(1) we fit the model III(BB+BB)*L3 to the PN-single data and calculate best-fit
parameters (2) we fold the model with variance in both the continuum and
line parameters
through MOS1 detector response and
simulate a spectrum with the same exposure time and spectral bins as
our analysis of the real data. (3) we evaluate $\Delta\chi^2$ by
fitting a simulated spectrum with and without a 3rd line. We repeat this
procedure 1000 times and compute the distribution of
$\Delta\chi^2$. Figure \ref{fig_compare_pns} shows the results from the
consistency
test between PN and MOS1 (left) and PN and MOS2 (right).

We calculated the fraction of simulated $\Delta\chi^2$ points smaller than
or equal to the 
measured $\Delta\chi^2$ ($S$), i.e. $f(\Delta\chi^2 \le S)$. The results are
   $f(\Delta\chi^2 \le S) = 2.9\times10^{-2}$ (MOS1) and
$f(\Delta\chi^2 \le S)= 3.1\times10^{-2}$ (MOS2). The PN and MOS1 results are
 just consistent at the
$1.90\sigma$ level and at the $1.86\sigma$ level between PN and MOS2.
We conclude that the behavior of PN and MOS instruments is internally
consistent
with regards fitting of the lines, as we showed previously was the case for the
continuum fits.


\subsection{Strength of the residuals in the Au--M region}

There are no statistically significant additional spectral features
in the 1E1207 spectrum. For thoroughness we indicate just how weak
the residuals are that we have uncovered at the Au-M edge. We do this
by considering the residuals relative to 3C273, where the same instrument
residuals are well-established \citep{kirsch04}. 
 
\begin{figure*}[ht]
\epsscale{0.8}
\plotone{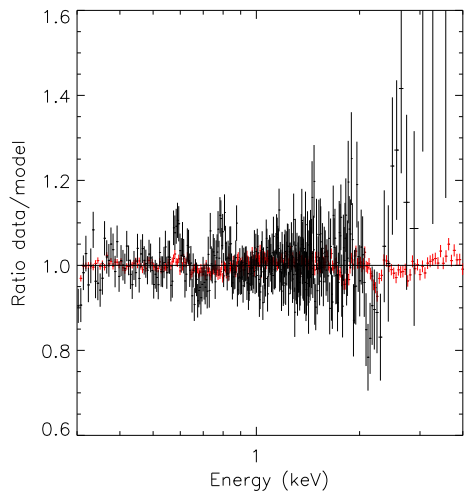}
\caption{Data/model ratio plot for 1E1207 (black) and 3c273 (red). \label{fig_3c273}}
\end{figure*}

We extracted PATTERN 0--4 events from a 40" radius region centered on
the source.  We used a double power law continuum model to fit the PN
singles data in the 0.3--6.0 keV energy range. We did not fit
the Au--M region but instead fit our model III with blackbody spectra for \pulsar\ and double power-law continuum for 3C273, and plot the ratio of the residuals to the model 
spectrum (figure \ref{fig_3c273}). We note that the residuals are $\sim$ 6--7\% in the quasar data and
$\sim$ 10--15\% in the neutron star data. The neutron star data has larger
error bars.

We directly calculated the EW of the 2.2 keV residuals without a line
fit, and find them to be $\sim$ 40 eV. This is consistent with the EW we determined by
line fitting. A similar calculation (as a sanity check) gave us 8 eV for the EW
residuals at 2.2 keV in 3C273, in agreement with measurements of the \xmm\  
team \citep{kirsch04}. Of course the much higher noise level in the 1E1207 data renders
the residuals insignificant. 
In figure 8 we show the ratio between the continuum model and the data for 1E1207. The same plot is shown for 3C273. We see there is nothing extraordinary in the NS data. The fluctuations
are uniform through the entire spectrum. In 3C273 the Au-M edge is readily
apparent because of the better counting statistics.

We note that a common error in estimating residual strength is to fit 
absorption lines and a continuum, and then to use the resultant continuum
without lines as
a basis for comparison with the residuals. This procedure grossly overestimates
the continuum, and thus the strength of any absorption features with respect to it.
The correct procedure of comparing the dips with respect to the best fit
continuum markedly reduces the significance of the residuals. With this proper
procedure our results here are consistent with those we obtained in \S 4.

\section{Conclusions \label{sec_conclusion}}

\begin{itemize}

\item Utilizing a wider range of physically plausible INS models and
careful background subtraction we have obtained continuum fits for 
1E1207 which effectively rule out the need for additional harmonics of
the first two spectral absorption features. Our derived continuum parameters
for both PN and MOS data are consistent with those found by previous
investigators.
 
\item To explore the significance of the weak residuals found at higher
energies in a few cases, we have implemented an improved F-statistic analysis
for additional weak lines in 1E1207. It corrects for systematic overestimation
of absorption line strengths using the standard F-test, as previously reported in
the literature. Calculating the correct, finite statistics reference distribution
for the F-statistic we find lines at harmonics of the two firmly established
lines are statistically insignificant.

\item We performed three completely independent statistical analyses to test for
the presence of third and fourth spectral features.  All tests give consistent results
and all tests indicate there is no third or fourth line.

\item While there are no statistically significant spectral features at 
higher energies, we explicitly fit the spectral residuals around 2.1 keV. Using
the continuum model favored by previous investigators, and for which we get
excellent continuum fits, we find the PN single spectrum residuals occur
at the position of the instrumental Au-M edge. The residuals appear 
at the same energy as those obtained by fitting the known Au-M
residuals in 3C273 with the identical procedure. We further show that 
PN doubles events yield a much lower energy for the residuals, as does the
combined PN singles/doubles data set. This is highly suggestive of an 
instrumental origin for the insignificant residuals. Indeed in all eight
physically plausible INS models we considered, the residuals always appear
at the Au-M edge.

\item We have demonstrated that the strength of the weak instrumental
residuals in the 1E1207 spectrum are consistent with the stronger 
instrumental residuals observed in the 3C273 spectrum.

\end{itemize}

\acknowledgements{We thank the referee for a very careful reading of the manuscript and for
suggesting changes which strengthened our analysis. We also thank the editor for numerous
suggestions of style, and for helping make the paper more concise and focused. JC acknowledges financial support from NASA's Graduate Student Researchers Program NGT5-50392. The authors thank Maurice Leutenegger for assistance and advice on various data processing issues.
 } 

\begin{appendix} 

\section{Testing for a spectral line in a continuum} 

Some important results used in \S\ref{sec_eadie} are summarized here. The
arguments of \citet{eadie} are closely followed.

Consider searching a region of interest (ROI) consisting of $K$ bins
for a deficit of counts over the expected continuum. The total 
spectrum has $N$ bins. Define $b_i$ as the true continuum, $\hat{b_i}$
as the estimated continuum, and $n_i$ as the observed counts in bin
$i$. We can define the difference of counts in bin $i$ as
$s_i=n_i-\hat{b_i}$. 

The search for residual in the $K$ bins of the ROI can be formulated
as a goodness-of-fit test, and under the null hypothesis $H_0$ there are no residuals, and $E(s_i)=E(n_i)-\hat{b_i} = b_i-\hat{b_i}\simeq
\hat{b_i}-\hat{b_i}=0$, where $E$ is the expectation (mean)
operator. A reasonable test
statistic is $T=\sum_{i=1}^K
\frac{(n_i-\hat{b_i})^2}{V(n_i-\hat{b_i})}$ where $V()$ is the
variance operator. This statistic is a function of the random
variables $z_i=\frac{n_i-\hat{b_i}}{\sqrt{V(n_i-\hat{b_i})}}$.

The $z_i$ have the following properties if $H_0$ is true. $E(z_i)=0$,
and $V(n_i-\hat{b_i}) = V(n_i)+V(\hat{b_i})-cov(n_i, \hat{b_i}) =
V(b_i)+V(\hat{b_i}) = \hat{b_i} + \hat{\sigma_i}^2$. $cov()$ is the
covariance operator. The trick is that the $\hat{b_i}$
should be estimated
using the $N-K$ bins outside the ROI, for then $n_i$ and $\hat{b_i}$ are
uncorrelated and their covariance vanishes. The continuum is simply
fit without utilizing the $K$ bins in the ROI. Also note that
$V(\hat{b_i}) \equiv \hat{\sigma_i}^2$ can be directly obtained from
covariance matrix of the continuum fit. Thus under $H_0$,
\begin{equation}
T=\sum_{i=1}^K \frac{(n_i-\hat{b_i})^2}{\hat{b_i}+\hat{\sigma_i}^2} =
\sum_{i=1}^K z_i^2
\end{equation}
and $E(z_i)=0, V(z_i)=1$. Thus the $z_i$ are 
standard, normal variables (for $s_i$ not too small), and $P(T)$ is
a $\chi^2$ distribution with $K$ degrees of freedom.

For $K \uax 30$ we can simplify this result by defining a new statistic
$y=\frac{T-K}{\sqrt{2K}}$. Since the mean and variance of the
$\chi^2$ distribution are $K$ and $2K$ respectively, $y$ has a standard,
normal distribution. Then as described in the text we can test $H_0$ 
by simply calculating $y$ and if $y< y_c=4$ we accept $H_0$.
\end{appendix}


\begin{deluxetable}{cccccccc}
\tablewidth{0pt}
\tablecaption{Best-fit continuum parameters to PN data \label{tab_pn_fit}}
\tablehead{ & PN & PN & PN & PN & PN & PN \\
  & I(BB+BB) & II(BB+BB) & III(BB+BB) & I(HA+HA)  & II(HA+HA) & III(HA+HA)}
\startdata
$N_H$ [$10^{22}$ cm$^{2}$]                      & $0.126_{  -0.017}^{ 
+0.009}$ & $0.128_{  -0.005}^{  +0.017}$         & $0.132_{  -0.014}^{+0.012}$ 
& $0.146_{-0.009}^{+0.009}$     & $0.151_{-0.010}^{+0.010}$  & $0.155_{-0.016}^{+0.007}$       \\
$kT^{\infty}_S$ [keV]                           & $0.159_{  -0.005}^{+0.009}$
   & $0.154_{  -0.017}^{  +0.010}$ & $0.153_{  -0.009}^{+0.015}$ 
& $0.116_{-
0.009}^{+0.009}$     & $0.109_{-0.006}^{+0.006}$ & 
$0.116_{-0.007}^{+0.009}$           \\
$R^{\infty}_S$ [km] \tablenotemark{a}                           & $5.01_{ 
-0.36}^{  +0.88}$            & $5.3_{  -0.5}^{  +1.3}$             & $5.51_{ 
-0.87}^{  +0.72}
$            & $21.2_{-3.7}^{+1.9}$       & $23.7_{-2.5}^{+3.5}$   & 
$21.2_{-3.9}^{+1.4}$            \\
$kT^{\infty}_H$ [keV]                           & $0.296_{  -0.012}^{ 
+0.010}$ & $0.287_{  -0.020}^{  +0.018}$ & $0.286_{  -0.010}^{  +0.024}$  & 
$0.358_{-0.057}^{+0.074}$
      & $0.316_{-0.040}^{+0.017}$  & $0.360_{-0.064}^{+0.060}$ 
\\
$R^{\infty}_H$ [km] \tablenotemark{a}                           & $0.97_{ 
-0.12}^{
+0.13}$   & $1.09_{  -0.23}^{  +0.19}$   & $1.10_{  -0.28}^{  +0.19}$ 
& $
0.68_{-0.28}^{+0.40}$        & $1.01_{-0.24}^{+0.54}$    & 
$0.67_{-0.23}^{+0.72}$              \\
$E_1$ [keV]                                             & $0.741_{ 
-0.006}^{  +0.008}
$         & $0.743_{  -0.007}^{  +0.007}$ & $0.725_{  -0.005}^{  +0.006}$ 
& $0.730_{-0.005}^{+0.005}$      & $0.731_{-0.005}^{+0.005}$ & 
$0.728_{-0.005}^{+0.005}$            \\
$\tau_1$              & $0.92_{  -0.07}^{  +0.05}
$         & $0.99_{  -0.10}^{  +0.26}$   & $0.60_{  -0.03}^{  +0.06}$  & 
$0.56_{-0.02}^{+0.02}$        & $0.58_{-0.01}^{+0.02}$   & $0.53_{-0.03}^{+0.02}$ 
\\
$FWHM$ [keV]                                     & $0.303_{  -0.026}^{  +0.024}$
   & $0.297_{  -0.022}^{  +0.015}$                 & $0.290_{  -0.020}^{  +0.030}$ 
& $
0.272_{-0.018}^{+0.023}$             & $0.262_{-0.016}^{+0.017}$    & $0.264_{-0.019}^{+0.020}$ 
\\
$E_2$ [keV]                                     & $1.398_{ -0.010}^{ 
+0.009}$
   & $1.382_{  -0.009}^{  +0.008}$  & $1.388_{  -0.008}^{  +0.008}$   & 
$1.391_{-0.008}^{+0.008}$      & $1.379_{-0.008}^{+0.008}$  & $1.389_{-0.009}^{+0.008}$ 
\\
$\tau_2$                                                & $2.9_{-1.2} \tablenotemark{b}$
         & $0.58_{  -0.10}^{  +0.16}$            & $0.39_{ 
-0.02}^{  +0.04}$ & $0.55_{-0.07}^{+0.05}$        & $3.7_{-2.0}^{+13}$    & 
$0.39_{-0.04}^{+0.04}$            \\
$FWHM$ [keV]           & $0.15_{ -0.10}^{+0.02}$   & $0.183_{  -0.013}^{  +0.026}$         & $0.187_{  -0.025}^{  +0.026}$ & $0.190_{-0.024}^{+0.027}$             & $0.140_{-0.052}^{+0.029}$      & $0.187_{-0.023}^{+0.028}$ 
\\
$\chi^2/dof$    & $0.98$ & $0.98$ & $0.99$ & $0.98$ & $0.98$ & $0.98$ 
\\
$dof$                   & $422$ & $422$ & $422$ & $422$ & $422$ & $422$ 
\\
\enddata
\tablecomments{For hydrogen atmosphere models, we fixed neutron star mass 
and radius to 1.4 $M_\odot$ and 10 km. $kT^\infty$ and $R^\infty$ are insensitive to 
variation of
  $M$ and $R$ \citep{zavlin98_2}.}
\tablecomments{All parameters include a roundoff digit except when it would have required displaying a fourth digit after the decimal point.} 
\tablenotetext{a}{For apparent radius, we assumed distance is 2.1 kpc and the error bars do not include uncertainty in distance. The distance to the supernova remnant PKS 1209-51 was measured to be $\sim 2.1$ kpc \citep{giacani00}.}
\tablenotetext{b}{No upper bound was obtained (see \S 3.4)} 
\end{deluxetable}

\begin{deluxetable}{cccccc}
\tablewidth{0pt}
\tablecaption{Best-fit continuum parameters to MOS data 
\label{tab_mos_fit}}
\tablehead{Instrument  & MOS1 & MOS1 & MOS2 & MOS2        \\
Model           & III (BB+BB) & III(HA+HA) & III(BB+BB) & III(HA+HA)}
\startdata
$N_H$ [$10^{22}$ cm$^{2}$]                      & 
$0.05_{-0.04}^{+0.10}$      &  $0.176_{-0.037}^{  +0.035}$    &   $0.075_{-0.053}^{+0.074}$ 
& $0.182_{-0.050}^{+0.011}$     \\
$kT^{\infty}_S$ [keV]                           & 
$0.197_{-0.027}^{+0.026}$       &
    $0.117_{-0.016}^{  +0.025}$ &         $0.204_{-0.030}^{+0.021}$ 
& $
0.123_{-0.020}^{+0.018}$     \\
$R^{\infty}_S$ [km]                             & $3.4_{-1.2}^{+1.5}$ 
&
     $20.3_{-7.7}^{+9.8}$  &           $3.0_{-0.8}^{+1.6}$         &
$18_{-6}^{+19}$        \\
$kT^{\infty}_H$ [keV]                           & 
$0.374_{-0.040}^{+0.074}$       &
     $0.41_{-0.06}^{+0.11}$ &        $0.373_{-0.039}^{+0.056}$ 
&
$0.418_{-0.063}^{+0.083}$     \\
$R^{\infty}_H$ [km]                             & $0.39_{-0.17}^{+0.27}$ 
&
    $0.49_{-0.20}^{+0.46}$ &             $0.37_{-0.17}^{+0.38}$         &
$0.43_{-0.17}^{+0.59}$        \\
$E_1$ [keV]                                             &  $0.700_{-0.031}^{+0.033}$ &
             $0.728_{-0.016}^{+0.013}$&    $0.718_{-0.046}^{+0.018}$ 
& $0.729_{-0.016}^{+0.013}$     \\
$\tau_1$   & $0.90_{-0.22}^{+0.52}$  &
             $0.62_{-0.05}^{+0.12}$ &    $0.68_{-0.12}^{+0.49}$ 
& $0.56_{-0.06}^{+0.09}$        \\
$FWHM$ [keV]                                     & $0.43_{-0.08}^{+0.22}$ 
&
     $0.330_{-0.031}^{+0.043}$   &                $0.39_{-0.06}^{+0.15}$ 
&
$0.319_{-0.037}^{+0.056}$             \\
$E_2$ [keV]                                     & $1.418_{-0.010}^{+0.009}$ 
&
     $1.417_{-0.009}^{+0.009}$ &       $1.403_{-0.009}^{+0.010}$ 
&
$1.402_{-0.009}^{+0.009}$      \\
$\tau_2$                                                & 
$0.46_{-0.05}^{+0.04} $  &
             $0.39_{-0.05}^{+0.05}$ &      $0.47_{-0.05}^{+0.05}$         &
$0.44_{-0.04}^{+0.04}$        \\
$FWHM$ [keV]                                             & 
$0.236_{-0.038}^{+0.043}$
           & $0.181_{-0.030}^{+0.035}$      &      $0.229_{-0.037}^{+0.035}$         &
$0.215_{-0.031}^{+0.034}$      &        \\
$\chi^2/dof$  & $0.85$ & $0.80$  & $1.20$   & $1.18$   \\
$dof$         & $139$  & $139$   & $137$    & $137$    \\
\enddata
\tablecomments{All parameters include a roundoff digit except when it would have required displaying a fourth digit after the decimal point.} 
\end{deluxetable}

\begin{deluxetable}{cccccc}
\tablewidth{0pt}
\tablecaption{Goodness-of-fit parameters of 3rd line test \label{tab_eadie}}
\tablehead{Continuum model & Parameters & PN single & PN double & MOS1 & MOS2 }
\startdata
 & $T_c$ & 112.8 & 71.38& 65.26& 65.26 \\ 
III(BB+BB) & $T$ & 85.90 & 43.42 & 24.79 & 41.03 \\ 
 & $1-C$ & $1.7\times10^{-2}$  & $5.4\times10^{-2}$ & $5.3\times10^{-1}$
 &  $ 3.1\times10^{-2}$\\
\hline
 & $T_c$ & 112.8 & 71.38 & 65.26 & 65.26\\ 
III(HA+HA) & $T$ & 80.45 & 42.62 & 20.47 &  34.68 \\ 
  & $1-C$  & $3.3\times10^{-2}$ & $6.3\times10^{-2}$  & $7.7\times10^{-1
}$  &  $1.2\times10^{-1}$ \\

\enddata
\tablecomments{Model I and II show very similar results as model III.}
\end{deluxetable}

\begin{deluxetable}{cccccc}
\tablewidth{0pt}
\tablecaption{Summary statistics for 3rd feature significance \label{tab_master}}
\tablehead{Instrument & Continuum model & $P(F)$ analysis \tablenotemark{a} & Goodness-of-fit test \tablenotemark{a, b} & EW analysis \tablenotemark{a,b}}
\startdata
 & I (BB*L3+BB)  & 1.77 & 2.13 & *            \\
&I (BB+BB*L3)   &  3.29  & 2.13  & 2.07  \\
& I (HA*L3+HA)  & 2.81  & 1.18        & 1.69  \\
& I (HA+HA*L3)  & 2.84 & 1.18   &       1.96    \\
PN-single& II (BB+BB*L3) & 3.27 & 2.22  & 2.18  \\
& II (HA+HA*L3) & 2.55         & 1.69  & 1.85 \\
& III (BB+BB)*L3 & 3.26 &  2.17 &  2.14  \\
& III (HA+HA)*L3 & 3.12    & 1.83        & 1.88          \\
\hline
& I (BB*L3+BB)  & 0.70   & 0.33 &  *            \\
&I (BB+BB*L3)   &  2.09   & 0.33  &    1.92  \\
& I (HA*L3+HA)  & 0.32      & -0.92      &  \#   \\
& I (HA+HA*L3)  & 0.32  & -0.92 &         \#     \\
MOS1 & II (BB+BB*L3) & 2.36   & 0.28          &  1.82    \\
& II (HA+HA*L3) & 1.02 & -0.83          & \# \\
& III (BB+BB)*L3 & 2.55 & -0.07 & 1.36  \\
& III (HA+HA)*L3 & 1.32 & -0.73        &   \#   \\
\hline
& I (BB*L3+BB)  & 0.69  & 1.78 &        *       \\
&I (BB+BB*L3)   &  2.21  & 1.78 &  1.65    \\
& I (HA*L3+HA)  &  0.27   & 1.05      & \#    \\
& I (HA+HA*L3)  & -1.43  & 1.05        &  \#            \\
MOS2 & II (BB+BB*L3) & 2.17    & 1.97 & 1.68      \\
& II (HA+HA*L3) & 1.14   & 1.37  & \# \\
& III (BB+BB)*L3 & 2.14  & 1.87  & 1.87   \\
& III (HA+HA)*L3 &  1.11 & 1.18 &   \#    \\
\enddata
\tablenotetext{a}{The units of significance are sigma.}
\tablenotetext{b}{These results have not been corrected for blind search. This will reduce the significance of the results (see \S 4).} 
\tablenotetext{*}{Low temperature continuum too weak to determine EW (see \S3.4 and \S4.3)}
\tablenotetext{\#}{The upper limit is 1-sigma.} 
\end{deluxetable}

\end{document}